\documentclass[%
 reprint,
 superscriptaddress,
 amsmath,amssymb,
 aps,
 prl,
]{revtex4-2}

\usepackage{graphicx}
\usepackage{dcolumn}
\usepackage{bm}
\usepackage{natbib}
\usepackage{bibentry}
\usepackage{hyperref}

\begin{document}

\title{Fano Resonance between Stokes and Anti-Stokes Brillouin Scattering}

\author{KwanTo Lai}
\affiliation{Department of Applied Physics, The Hong Kong Polytechnic University, Hong Kong SAR}

\author{Daniel Finkelstein-Shapiro}
\affiliation{Division of Chemical Physics and NanoLund, Lund University, Sweden}

\author{Arnaud Devos}
\affiliation{Institut d’Electronique, de Microélectronique et de Nanotechnologie, Unité Mixte de Recherche, CNRS 8250, Villeneuve d’Ascq Cedex, France}

\author{Pierre-Adrien Mante}
\email{pierre-adrien.mante@chemphys.lu.se}
\affiliation{Department of Applied Physics, The Hong Kong Polytechnic University, Hong Kong SAR}
\affiliation{Division of Chemical Physics and NanoLund, Lund University, Sweden}

\date{\today}

\begin{abstract}

In recent years, the manipulation of Fano resonances in the time domain has unlocked deep insights into a broad spectrum of systems' coherent dynamics. Here, inelastic scattering of light with coherent acoustic phonons is harnessed to achieve complex Fano resonances. The sudden change of phonon momentum during reflection leads to a transition from anti-Stokes to Stokes light scattering, producing two different resonances that interfere in the measurement process. We highlight the conditions necessary to achieve such interference, revealing an underlying symmetry between photons and phonons, and verify the theory experimentally. Then, we demonstrate the possibility to characterize energy and coherence losses at rough interfaces, thus providing a mechanism for non-destructive testing of interface quality. Our results describe numerous unexplained observations in ultrafast acoustics and can be generalized to the scattering of light with any waves.

\end{abstract}

\maketitle

Fano resonances occur when a broad and featureless continuum of states couples to a discrete resonance with energy lying within the continuum, thereby opening two interfering pathways to reach the continuum manifold from an initial state. This results in an asymmetric lineshape in the system's frequency response, characterized by the Fano parameter, $q$\cite{Fano1961}. Fano resonances were first studied in the context of photoionization~\cite{Beutler1935} but have since been found in the optical response of metamaterials and plasmonic nanostructures~\cite{Lucky2010,Miroshnichenko2010,Gallinet2010}, and also in mechanical~\cite{PhysRevB.101.024101} and electronic systems~\cite{Xiao2016}. Interference-based sensors rely on using frequency shifts, more easily measured than amplitude modulations. Therefore, Fano resonances are extremely sensitive to perturbations and have been used for characterization and sensing~\cite{Lucky2010, Xiao2016,doi:10.1021/nl200135r}.  Furthermore, Fano resonances can also be used to investigate dissipative mechanisms ~\cite{PhysRevLett.86.4636,Baernthaler2010}. The presence of relaxation breaks the time-reversal symmetry and can be described by a complex Fano parameter~\cite{PhysRevLett.86.4636,PhysRevA.97.023411,PhysRevA.93.053837}. By analyzing the trajectories of the $q$ parameter in the complex plane as a function of dissipation strength, dissipative (energy loss) vs. decoherence (coherence loss) were succesfully distinguished~\cite{Baernthaler2010}.

Fano resonances have also been investigated in the time-domain, allowing additional characterization~\cite{PhysRevLett.74.470,PhysRevLett.94.023002} and deeper insights into their properties~\cite{Ott716, Kaldun738}. Notably, such studies have shown the coherent nature of magnetoexcitons in GaAs~\cite{PhysRevLett.74.470} and of electronic wavepackets~\cite{PhysRevLett.94.023002}. In the time domain, the Fano resonance appears as the interference between a long lived state (the discrete state) and a short lived state (akin to the continuum). These time-domain investigations of Fano resonances have showcased the mapping of the Fano parameters, $q$ in the frequency domain to a phase, $\phi$ in the time domain~\cite{Ott716}. A consequence of this relation is the possibility of manipulating Fano resonance using a perturbation that introduces a phase-shift in the response of the system's discrete pathway, even when a Fano structure is absent~\cite{Ott716}. Such concepts have been used to launch and interrupt a Fano resonance~\cite{Kaldun738} or to observe Fano resonance by inducing simultaneous stimulated emission and absorption~\cite{Kotur2016}. Up to now, manipulation of Fano resonances through ultrafast perturbation were limited to atomic systems, however, applying such concept to condensed matter would unlock vast opportunities for material characterization thanks to the extreme sensitivity of Fano resonance.  

In this letter, we investigate the formation of Fano resonances due to the interaction of light with coherent acoustic phonons (CAPs) reflecting at an interface. Although minima in the measured phonon spectrum had been routinely observed, neither the connection to Fano interferences nor an explanation of the underlying physical mechanism has been provided. We first briefly review the Stokes and anti-Stokes interaction of light with CAPs and highlight the transition from anti-Stokes to Stokes Brillouin scattering during the reflection of CAPs at an interface. During this transition, light undergoes both scattering simultaneously, leading to interference in the scattering probability, responsible for forming a complex Fano resonance. We perform picosecond ultrasonic experiments on a thin Tungsten (W) film that reveals a Fano resonance in the transient reflectivity spectrum in excellent agreement with our model. We finally discuss the possibility of dissociating dissipation from decoherence by tracking the $q$ parameter in the complex plane as a function of the laser wavelength, thus enabling non-destructive characterization of the roughness of surfaces and buried interfaces. 

In condensed media, Brillouin scattering refers to the scattering of light by waves that modifies the material's refractive index, such as phonons~\cite{refId0}, magnons~\cite{BOROVIKROMANOV1982351}, or polaritons~\cite{YU197929}. With coherence, additional effects can occur due to the wave well-defined phase and momentum, like parametric interactions~\cite{Otterstrom1113, Bahl2012} or the possibility to perform time-resolved investigations~\cite{PhysRevB.81.113305, PhysRevLett.111.225901,PhysRevB.34.4129, PhysRevLett.86.2669, he_acoustic_2017, MATSUDA20153, PhysRevLett.103.264301}. In the following, we show how we take advantage of the waves' coherent nature to create a Fano response in the scattering of light. We take the example of CAPs, but the results can be generalized to any type of light scattering waves.  
\begin{figure}
    \centering
    \includegraphics[width=8.5 cm]{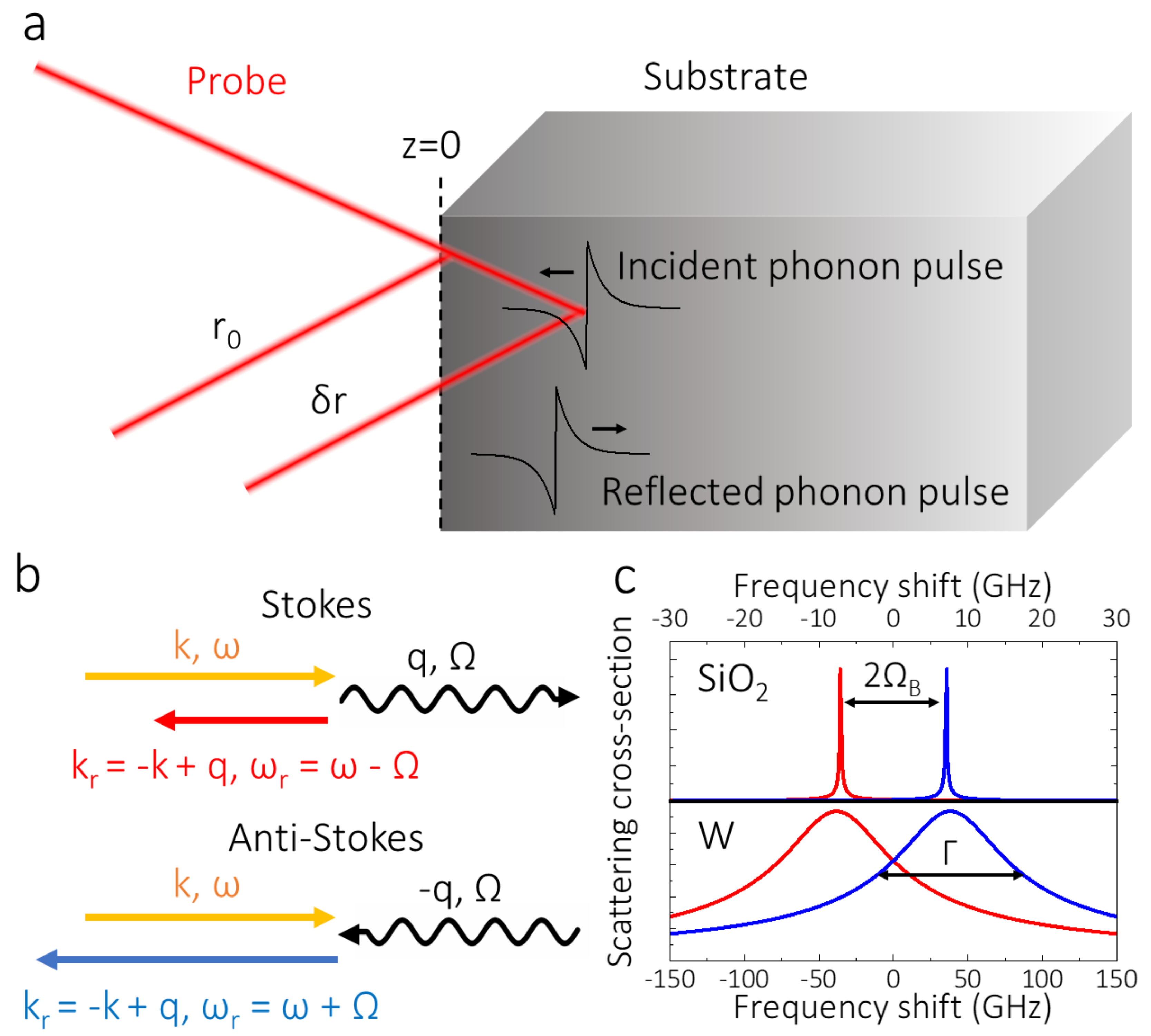}
    \caption{\textbf{Principles of light scattering by coherent acoustic phonons}. (a) A femtosecond laser (probe) is reflected by a free surface and by a coherent phonon pulse propagating within a semi-infinite substrate and getting reflected at the free surface. (b) Energy and momentum conservation rules before and after the reflection of the phonon pulse. (c) Lineshape of Stokes (red) and anti-Stokes (blue) resonance in SiO$_2$ and W. When the material is transparent, the resonances are not overlapping and cannot interfere}
    \label{fig:principle}
\end{figure}

We first investigate the scattering of light by CAPs. We consider a picosecond ultrasonic experiment at normal incidence performed on an absorbing thin film in the region $z>0$ (Fig.~\ref{fig:principle}a). In this experiment, a femtosecond laser pulse, the pump, is absorbed by the material, leading to the generation of longitudinal CAPs, $\eta(z-vt)$, that propagates in the depth of the sample~\cite{PhysRevB.34.4129, PhysRevLett.86.2669}. A second time-delayed femtosecond laser pulse, the probe, is Stokes scattered by the propagating CAPs depicted in the upper part of Fig.~\ref{fig:principle}b. This scattering process corresponds to the emission of a phonon at the angular frequency $\Omega_B=4\pi nv/\lambda$ with $n$, the refractive index at the probe wavelength, $\lambda$, and $v$, the longitudinal sound velocity. The CAPs then propagate to the bottom of the thin film, are reflected towards the surface, and re-enter the probe's penetration region. This time, the probe light is anti-Stokes scattered, corresponding to the absorption of a phonon at the frequency $\Omega_B$, as shown in the lower part of Fig.~\ref{fig:principle}b. During the following reflection of the CAPs, the probe light simultaneously undergoes Stokes and Anti-Stokes process due to the change of phonon momentum. Under certain conditions, the probabilities of Stokes and anti-Stokes scattering may interfere, leading to the formation of a Fano resonance in the scattering cross-section. 

We now study the appearance of such Fano resonance. For simplicity in the following, we assume that the reflection of CAPs at the free surface occurs at $t=0$. The time-dependent relative reflectance, or scattering cross-section, for a monochromatic electromagnetic wave at frequency $\omega$, is given by (see Supplemental materials)~\cite{PhysRevB.34.4129,MATSUDA20153}:
\begin{multline}
    \frac{\delta r(t)}{r_0}=\rho e^{-j\phi}\int_{-\infty}^{+\infty}dz \eta(vt-z)\\
    \left[H(-z)e^{-2jk\tilde{n}z}+RH(z)e^{2jk\tilde{n}z}\right],
    \label{eq:Time_domain}
\end{multline}
where 
\begin{equation}
    \rho e^{-j\phi}=\frac{4jk\tilde{n}}{1-\tilde{n}^2}\left(\frac{\partial n}{\partial \eta}+j\frac{\partial \kappa}{\partial \eta}\right),
\end{equation}
and $k$ is the wavevector of the probe light, $R$ is the acoustic reflection coefficient at the free surface, $H(z)$ is the Heaviside function, $\tilde{n} = n+j\kappa$ and $\frac{\partial n}{\partial \eta}+j\frac{\partial \kappa}{\partial \eta}$ are the complex refractive index and complex photo-elastic coefficient at the probe wavelength, respectively. Eq.~\eqref{eq:Time_domain} is the convolution of the CAPs with the Green's function of the system, \textit{i.e.} the light scattering cross-section for a $\delta$-like CAPs pulse. The first term in the square bracket corresponds to anti-Stokes scattering and the second to the Stokes process. At the surface ($z=0$), the Green's function is discontinuous, \textit{i.e.} the reflection of the CAPs introduce a phase shift.

The scattering cross-section $\sigma(\Omega,\omega)$ for photons of frequency $\omega$ by phonons of frequency $\Omega$ is obtained by performing the Fourier transform of Eq.~\eqref{eq:Time_domain} (see Supplemental materials):
\begin{equation}
    \sigma(\Omega,\omega)=\frac{\rho e^{-j\phi}}{\Gamma}\left[\frac{1+j\chi}{1+\chi^2}+R\frac{1-j\Delta}{1+\Delta^2}\right],
    \label{eq:FT}
\end{equation}
where we have introduced $\Gamma=2 \kappa \omega v/c$, $\chi=(\Omega+\Omega_B)/\Gamma$ and $\Delta=(\Omega-\Omega_B)/\Gamma$. The scattering cross-section of Eq.~\ref{eq:FT} is composed of two resonances corresponding to the Stokes and anti-Stokes scattering. The amplitude of each resonance taken separately is shown in Fig.~\ref{fig:principle}c for the case of SiO$_2$ and W. One notices, in Eq.~\ref{eq:FT}, that photons and phonons are the mirror of each other: looking at the evolution of $\sigma(\Omega,\omega)$ for a fixed $\omega$ and a varying $\Omega$, or vice versa, we obtain two resonances corresponding to the resonant Stokes and anti-Stokes scattering for photons, or the resonant absorption and emission for phonons.

Finally, we can rewrite Eq.~\ref{eq:FT} as the sum of a complex Lorentzian and a complex Fano resonance (see Supplemental materials): 
\begin{equation}
    \sigma(\Omega,\omega)=\frac{\rho e^{-j\phi}}{\Gamma(1-j\chi)}\bigg[\frac{(\Delta+q)^2+(1+jq)^2}{1+\Delta^2}\bigg]
    ,\label{eq:Fano}
\end{equation}
where $q=-R(j+\chi)/2$ is the Fano parameter. The Fano resonance is thus resulting from the interference between the Stokes and anti-Stokes scattering of light. 

By analyzing the Fano parameter, we can achieve deeper insights into the phenomenon. We see that the real part, $-R\chi/2$, is proportional to $n/\kappa$ in the vicinity of the resonance (see Supplemental materials) and represents the overlap between the Stokes and anti-Stokes resonances shown in Fig.~\ref{fig:principle}c. When $n$ increases, the Brillouin scattering frequency increases, reducing the overlap between Stokes and Anti-Stokes. This leads to a disappearance of the Fano resonance as $q \rightarrow \infty$. Moreover, when $\kappa$ increases, the penetration of light is smaller, which broadens Brillouin resonances and increases the overlap between Stokes and anti-Stokes resonances. In Fig.~\ref{fig:principle}c, we can see how the absorption impacts the overlap of resonance. In SiO$_2$, absorption is weak, and the resonances do not overlap, contrary to W. Finally, the reflection coefficient, $R$, modifies the overlap by reducing the amplitude of the Stokes resonance. The imaginary part of the Fano parameter reflects the energy lost during the reflection process through the acoustic reflection coefficient $R$.  

\begin{figure}
    \centering
    \includegraphics[width=8.5 cm]{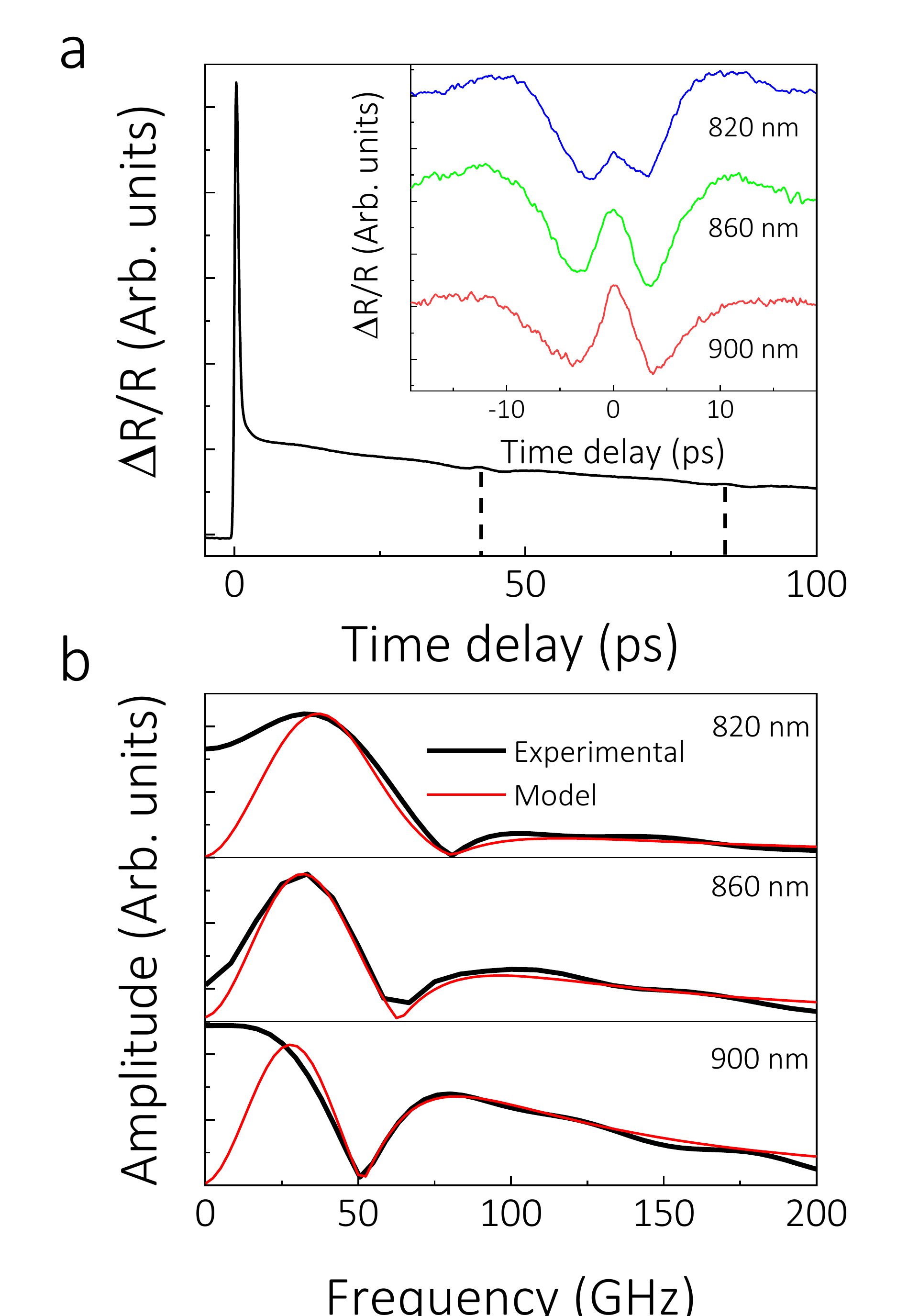}
    \caption{\textbf{Experimental investigation and modeling.} (a) Transient reflectivity obtained on a 110 nm W film deposited on a Si substrate for a pump and probe wavelength of 860~nm. Inset: echoes obtained for a probe wavelength of 820 (blue), 860 (green) and 900~nm (red). (b) Fourier transform of the acoustic echoes (black lines) obtained at 820~nm (upper panel), 860~nm (middle panel) and 900~nm (lower panel) and model (red lines)}
    \label{fig:W}
\end{figure}

We carried out picosecond ultrasonic experiments using a tunable Ti:sapphire oscillator and a conventional one-color pump and probe setup at normal incidence to investigate the appearance of Fano interferences between Stokes and anti-Stokes Brillouin scattering. The laser produces 120~fs optical pulses at a repetition rate of 80 MHz, centered at a wavelength tunable between 690 and 1040~nm. We performed these experiments on a 110~nm thick W thin film deposited on a Si substrate for three wavelengths: 820, 860, and 900~nm. The transient reflectivity obtained for a wavelength of 860~nm is represented in Fig.~\ref{fig:W}a. 

One first remarks an initial rise of the reflectivity produced by photogenerated carriers. A decay follows due to electron-electron and electron-phonon scattering. We also observe structures at 42 ps and 84~ps, called acoustic echoes. They correspond to the detection of the CAPs after one and two round-trips in the thin film, respectively. Considering a longitudinal sound velocity of 5200~m.s$^{-1}$ \cite{samsonov_handbook_1968}, we obtain a round-trip time of 42.3~ps. In the inset of Fig.~\ref{fig:W}a, a zoom on the first echoes obtained for different wavelengths is shown. We see an oscillation growing in amplitude until $t=0$, which corresponds to the CAPs propagating towards the surface. After $t=0$, the signal reverses due to the reflection of the CAPs and the following propagation towards the substrate. The echoes' shape is changing due to the variation of the optical and photo-elastic properties with the wavelength~\cite{PhysRevLett.86.2669}. 

We now model the interaction of light with these CAPs using our analysis. In picosecond ultrasonic experiments, we measure the transient reflectivity, which is given by $\frac{\Delta R}{R_0} \simeq 2Re(\frac{\delta r}{r_0})$. The spectrum of the transient reflectivity is thus given by the real part of the product of the scattering cross-section (Eq.~\ref{eq:Fano}) and the frequency content of the pulse. We consider the strain to be an odd function, which is a good approximations for CAPs generated at a free surface~\cite{PhysRevLett.86.2669, MATSUDA20153, PhysRevB.34.4129}. In that case, the spectrum of the strain is imaginary; the transient reflectivity spectrum is therefore proportional to the imaginary part of Eq.~\ref{eq:Fano}. Taking into account these considerations, we can calculate the spectra of these echoes for different wavelengths (see Supplemental materials).

In Fig.~\ref{fig:W}b, we show the Fourier transform for the echoes obtained at 820, 860, and 900~nm. In each case, we observe a Fano-type lineshape with a dip in the spectrum corresponding to the destructive interferences. When the wavelength increases, the Fano lineshape transforms: the frequency of the Fano minimum decreases and gets closer to the Fano maximum, corresponding to a decrease of the Fano parameters, in agreement with the ratio $n/\kappa$ that varies from 1.22 at 820~nm to 1.08 at 900~nm. We also reproduce the spectra obtained using our model (see Supplemental materials). Here, we used the refractive index values given in~\cite{Rakic:98} and an acoustic reflection coefficient of -1, corresponding to a perfect reflection of the CAPs. For the photo-elastic coefficients, we used values from~\cite{PhysRevLett.86.2669}. We observe an excellent agreement between our model and experimental data confirming that the transient reflectivity induced by the reflection of CAPs pulse can be modeled as a complex Fano resonance between the Stokes and anti-Stokes Brillouin scattering amplitudes. The differences at low frequencies come from the subtraction of the electronic contribution to the signal. Previous investigations observed such a dip in the acoustic echoes' spectrum~\cite{he_acoustic_2017,MATSUDA20153, PhysRevB.81.113305}, but their origin has not yet been connected to Fano resonances. We have applied our model to these observations and can reproduce these results (see Supplemental Materials), showing how general this phenomenon is.

\begin{figure}
    \centering
    \includegraphics[width=8.5 cm]{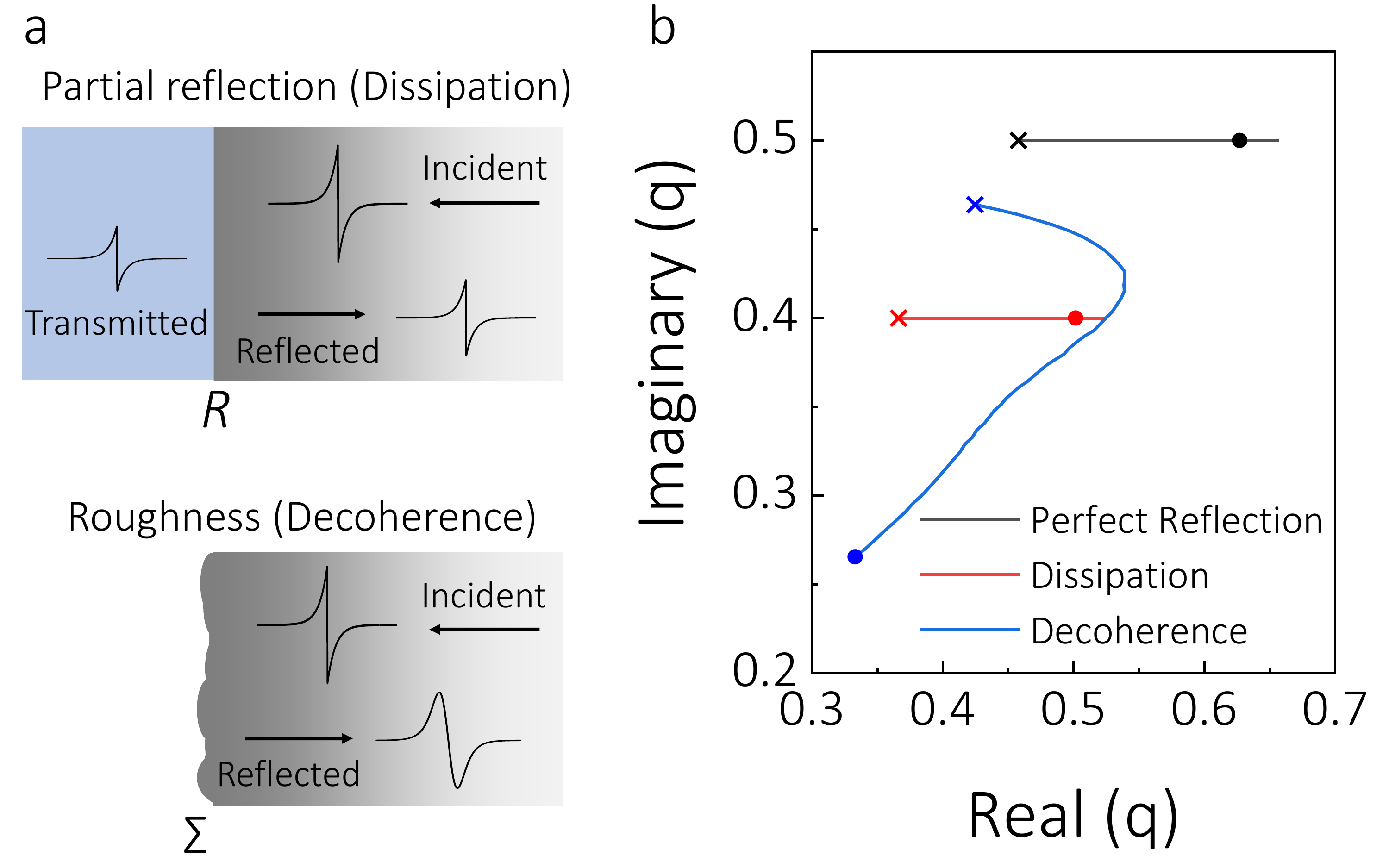}
    \caption{\textbf{Dissipation vs. Decoherence} (a) The two scenarii considered corresponding to dissipation due to partial reflection at an interface characterized by the reflection coefficient $R$, and to decoherence due to the surface roughness, characterized by the standard deviation of roughness $\Sigma$. (b) Evolution of the real and imaginary parts of the Fano parameter for wavelengths ranging from 350 to 1000~nm for the case of a perfect reflection (black line), dissipation (red line) and decoherence (blue line). The beginning and ending of the wavelength range are marked by a cross and circle, respectively.}
    \label{fig:dissipation v. decoherence}
\end{figure}

We have demonstrated the possibility to observe Fano resonances due to the interference between the parametric emission and absorption of phonons by light. Now, we show how such engineered Fano resonance can be used to perform interface characterization. Previous work has shown that an analysis of the $q$ parameter trajectory in the complex plane reveals the origin of incoherent dynamics~\cite{Baernthaler2010}. Here, we model dissipation and dephasing by introducing a modified reflection coefficient, $\tilde{R}=R e^{-8k^2n^2\Sigma}$, where $\Sigma$ is the standard deviation of the normally distributed phase added to the CAPs upon reflection at the surface (see Fig.~\ref{fig:dissipation v. decoherence}a and Supplemental materials). The roughness of the interface can produce such decoherence of the acoustic strain~\cite{PhysRevLett.111.225901,PhysRevLett.103.264301}. We investigate the trajectory of the Fano parameter in the wavelength range 350 to 1000~nm (See Fig.~\ref{fig:dissipation v. decoherence}b). We examine the case
of perfect reflection ($R$ = -1, $\Sigma$ = 0~nm), of dissipation ($R$ = -0.8, $\Sigma$ = 0~nm) and of decoherence (R = -1, $\Sigma $ = 10~nm).

We observe a striking difference between the trend for dissipation and decoherence. In the case of dissipation, only the real part, given by $Rn/\kappa$, is changing since the imaginary part, which represents losses, is dictated by the acoustic reflection coefficient, considered constant regardless of the phonon frequency. On the other hand, we observe that the Fano parameters' real and imaginary parts are simultaneously changing for decoherence. Decoherence impacts CAPs differently as a function of their wavelength, as can be seen in the broadening of the reflected CAPs in Fig.~\ref{fig:dissipation v. decoherence}a. When changing the probe wavelength, CAPs at different frequencies are resonantly probed, and the imaginary part of the Fano parameter is more or less impacted. The roughness of a surface induces diffusive scattering of CAPs that destroy their coherence~\cite{PhysRevLett.103.264301,PhysRevLett.111.225901}. Currently, interface roughness is obtained by comparing the loss in amplitude of signal produced by CAPs before and after reflection for different frequency~\cite{PhysRevLett.103.264301}. Fano resonance sensitivity could be used to achieve more precise and reliable non-destructive measurements of the roughness of surfaces and buried interfaces.

In conclusion, we predict and model the complex Fano resonances resulting from interference in the scattering amplitude of light by coherent acoustic phonons reflecting at an interface. We highlight the conditions necessary for the appearance of such lineshapes. We then experimentally verify their appearance by performing picosecond ultrasonic experiments on a thin W film. We obtain an excellent agreement between the experimental observations and our model. We also apply our model to other reports from literature, highlighting the universality of this phenomenon. Finally, we demonstrate how we can dissociate dissipation from decoherence by monitoring the evolution of the complex Fano parameter as a function of wavelength. Our results can be extended, not only to other types of propagating waves that scatter light but more generally to the interference between parametric emission and absorption. The observed Fano resonance and the enhanced sensitivity of such interferometric methods open the way for improved and novel characterizations, particularly for interface characterization and interfacial thermal transport in the case of phonons.

This work was supported by NanoLund, the Crafoord Foundation, and Grant 2017-05150 from the Swedish Research Council (VR).

\bibliographystyle{apsrev4-2}
\bibliography{biblio}

\end{document}